\documentstyle[cite,12pt]{article}
\newcommand{\be}{\begin{equation}}
\newcommand{\ee}{\end{equation}}
\newcommand{\beqn}{\begin{eqnarray}}
\newcommand{\eeqn}{\end{eqnarray}}
\newcommand{\beqnn}{\begin{eqnarray*}}
\newcommand{\eeqnn}{\end{eqnarray*}}

\def\s{\sigma}

\def\vep{\varepsilon}

\begin{document}

\title{Classicality and anticlassicality measures of pure and mixed
quantum states}

\author{
V.V. Dodonov \thanks{e-mail: vdodonov@df.ufscar.br}
\thanks{
on leave from Lebedev Physics Institute and Moscow Institute of Physics and
Technology, Russia}, \
 M.B. Ren\'o \thanks{e-mail: pmbr@df.ufscar.br} \\
Departamento de F\'{\i}sica, Universidade Federal de S\~ao Carlos,\\
Via Washington Luiz km 235, 13565-905 S\~ao Carlos, SP, Brazil
}

\date{}
\maketitle


\begin{abstract}
We introduce a simple measure of ``classicality'' of pure and mixed quantum
states as a maximum value of the Hilbert--Schmidt ``scalar products''
between the renormalized statistical operators of the state concerned and all
displaced thermal states. Choosing
Fock states as the reference set, we introduce the measure of
``anticlassicality''. Both measures are illustrated for
the Fock, coherent phase, and generic mixed Gaussian states.
Gaussian states are shown to be the closest to thermal states possessing
the same degree of quantum purity. On the contrary, Fock states appear to
be more close to mixed thermal states than to pure coherent states.
\end{abstract}

\vspace{5mm}



\newpage

\section{Introduction}

Since Glauber's paper \cite{Glauber},
quantum states for which the so-called $P$-distribution \cite{Sud1} is
nonpositive or more singular than delta function are called
``nonclassical''.
For the past decades, many authors proposed different {\em quantitative\/}
measures of ``nonclassicality''.
It seems impossible to reduce all variety of quantum states in
infinite-dimensional Hilbert spaces to some unique parameter.
Therefore, different existing approaches should be considered sooner
as complementary rather than competitive.

Historically, the first approach was based on the analysis of deviations
from the Poissonian photon statistics inherent to the
Klauder--Glauber--Sudarshan \cite{Glauber,Sud1,Klaud1}
{\em coherent states\/}, which are known to be the only ``classical''
{\em pure\/} states \cite{Ahar66,CarNiet,Hill85}.
The examples are Mandel's ``$Q$-parameter'' \cite{Mand-Q} and its
various generalizations \cite{Lee,AgT2,Klysh}.
The second direction is to evaluate the volume of that part of phase space
where some quasiprobability distribution assumes negative values
\cite{Wlod88,KazOr90,Lee91,LutBar,Jansz,Kim99,MarBas01,Paris,Yoon}
(see also \cite{Chum,Arv}). There are also other approaches
\cite{Vog,Hon01,Lvov02}.

Here we follow the direction opened by Hillery \cite{Hill}. It consists
in evaluating some kinds of {\em distances\/} in the Hilbert space
between the state concerned and a family of states
which are assumed to be ``classical''.
There exist many different definitions of the distance, besides
the ``trace distance'' used in \cite{Hill}.
In particular,
distances along geodesics on curved manifold supplied with
Riemannian metrics for coherent, squeezed, displaced, and other states
were studied in \cite{Provost,Trif,Abe}.
Other examples are the Monge distance \cite{Zyc},
``polarized'' and ``classical-like'' distances \cite{dw1}.
The most simple from the point of view of calculations is
the Hilbert--Schmidt distance, used in \cite{dw1,Dieks,Wun,Orl,Buz,dw2,DW3},
\begin{eqnarray}
d^2_{HS}(\hat{\rho},\hat{\rho}_c)
&=&{\rm Tr}(\hat{\rho}-\hat{\rho}_c)^2
\nonumber\\
&=&{\rm Tr}(\hat{\rho}^2)+
{\rm Tr}(\hat{\rho}_c^2) -2\,{\rm Tr}(\hat{\rho}\hat{\rho}_c).
\label{d-HS}
\end{eqnarray}
Here $\hat{\rho}$ is the statistical operator of the quantum state concerned,
and $\hat{\rho}_c$ is related to the reference ``classical'' state.

For {\em pure\/} quantum states, $\hat{\rho}=|\psi\rangle\langle\psi|$,
the reference states are usually identified with {\em coherent states\/}
$\hat{\rho}_c=|\alpha\rangle\langle\alpha|$, as soon as the latter
are assumed to be
``the most classical'' ones \cite{Ahar66,CarNiet,Hill85}.
Then the calculation of
distance is reduced to calculating the scalar product
$\langle\alpha|\psi\rangle$. This scalar product is the principal ingredient
of many other distances, such as the Fubiny--Study distance
\cite{Barg}
$d^2_{FS}=1-|\langle\psi_1|\psi_2\rangle|^2$
(we use here slightly different notation),
or Wootters' distance \cite{Woot,Braun}
$d_{W}=\cos^{-1}(|\langle\psi_2|\psi_1\rangle|)$
(see \cite{dw1} for more references).

However, coherent states (whose $P$-distributions are delta-functions)
represent only a small subset ``at the border'' of the set of all
``classical'' states.
Therefore, it seems reasonable to enlarge the family of reference states
$\hat\rho_c$, incorporating \cite{dw2} all {\em displaced thermal states\/},
in complete correspondence with original paper \cite{Glauber}.

Unfortunately, even the Hilbert--Schmidt distance becomes complicated
in such a case, because the presence of three terms
given in the second line of Eq. (\ref{d-HS}) does not permit to find
the minimum of this expression analytically (except for the most simple
cases), when one deals with {\em mixed\/} states.
A possibility to simplify calculations by adjusting the ``purity''
${\rm Tr}(\hat{\rho}_c^2)$ of the reference states to the purity
${\rm Tr}(\hat{\rho}^2)$ of the state concerned was discussed in \cite{dw2}.

The aim of our article is to consider another possibility, which
permits us to find analytical expressions for most known families of
nonclassical states. Namely, in the next section we introduce a
new ``classicality measure'', which is proportional to
({\em but not identical with\/}) the last term in (\ref{d-HS}),
${\rm Tr}(\hat{\rho}\hat{\rho}_c)$. We analyze this measure for the
Fock and generic Gaussian mixed states.
In Sec. 3 we consider the ``anticlassicality measure'', based on using
``the most nonclassical'' Fock states as the reference ones.
A brief discussion of results and perspectives is given in the concluding
Sec. 4

\section{Classicality measure}

As a matter of fact, practically all information about the closeness
of the states $\hat{\rho}$ and $\hat{\rho}_c$ is contained
in the term ${\rm Tr}(\hat{\rho}\hat{\rho}_c)$, whereas the presence
of the term ${\rm Tr}(\hat{\rho}_c^2)$ in (\ref{d-HS}) only complicates
the search of minimum of $d_{HS}^2$, adding no significant information.
Therefore, it seems
reasonable to consider only the term ${\rm Tr}(\hat{\rho}\hat{\rho}_c)$,
searching for its {\em maximum\/} with respect to the enlarged
family of ``classical'' states, consisting of all displaced thermal states
\cite{MolGla,Sem84,BiVour}
\be
\hat\rho_c = \hat{D}(\alpha)\hat\rho_{th}\hat{D}^{-1}(\alpha).
\label{dts}
\ee
Here
$
\hat{D}(\alpha)=\exp\left(\alpha \hat{a}^{\dagger} -\alpha^{*} \hat{a}
\right)
$
is the well-known displacement operator
\cite{Glauber,Sud1,Klaud1}
written in terms of the bosonic
ahhihilation and creation operators, $[\hat{a},\hat{a}^{\dagger}]=1$,
and $\hat\rho_{th}$ is the thermal state,
\be
\hat\rho_{th}= (1-\eta)\sum_{k=0}^{\infty} \eta^k |k\rangle\langle k|,
\quad 0\le \eta <1.
\label{rhoth}
\ee
If we confine ourselves to the special case of $\eta=0$, when
$\rho_c=|\alpha\rangle\langle\alpha|$, and consider only pure quantum
states $\hat{\rho}=|\psi\rangle\langle\psi|$, then the quantity
$\max_{\alpha}|\langle \alpha|\psi\rangle|^2$ seems a good measure
of ``classicality'', because it equals $1$ for coherent states and
is less than $1$ for all other pure states.
However, dealing with the generic case of mixed states, one meets
certain difficulties: on the one hand, the measure of classicality
for any thermal state must be the same as for coherent states, i.e.,
it must be equal to $1$ according to the normalization chosen,
but on the other hand, ${\rm Tr}(\hat{\rho}_c^2)<1$.
Therefore, one has to find a suitable generalization of the scalar product
between pure states to the case of mixed states, which would result in
the unit value for identical states.

One possible solution, satisfying many additional requirements,
was given by Uhlmann \cite{Uhlm},
who showed that the generalization of the quantity
$|\langle\psi_1|\psi_2\rangle|$ can be taken in the form
$\mbox{Tr}\sqrt{\hat{\rho}_1^{1/2}\hat{\rho}_2 \hat{\rho}_1^{1/2}}$.
However, the calculation of this trace is rather complicated problem,
although certain progress has been achieved recently
for Gaussian states (in the case of infinite-dimensional Hilbert space)
\cite{Twam,Wang,scut,MarMar02}.

We prefere to follow a more straightforward way:
simply to replace the operators $\hat{\rho}$ and $\hat{\rho}_c$ in
${\rm Tr}(\hat{\rho}\hat{\rho}_c)$ by the {\em renormalized\/} operators
$\hat{\rho}^{\prime}$ and $\hat{\rho}_c^{\prime}\,$, where
\be
\hat\rho^{\prime}\equiv \hat\rho
/\sqrt{\mbox{Tr}\left(\hat\rho^2\right)}\,.
\label{rhoprim}
\ee
Therefore, we define the ``classicality measure'' as
\be
{\cal C}=\max_{\rho_c}\mbox{Tr}\left(\hat\rho^{\prime}
\hat\rho_{c}^{\prime}\right).
\label{def-C}
\ee
For pure states, $\hat{\rho}=|\psi\rangle\langle\psi|$, we have
\be
{\cal C}=\max_{\rho_c}\langle\psi|\hat\rho_{c}^{\prime}|\psi\rangle.
\label{Cpur}
\ee
In modern literature, the ``generalized scalar products'', such as
$\mbox{Tr}\sqrt{\hat{\rho}_1^{1/2}\hat{\rho}_2 \hat{\rho}_1^{1/2}}$ or
${\rm Tr}(\hat{\rho}_1\hat{\rho}_2)$, are frequently called
\cite{Joz,Schum} ``fidelities''
(``Bures--Uhlmann'' or ``Hilbert--Schmidt'', respectively).
In particular, a possibility of using the ``modified fidelity''
${\rm Tr}(\hat{\rho}_1^{\prime}\hat{\rho}_2^{\prime})$ was mentioned
in \cite{Duan}.
Note that if only one of the states is pure, then the ``modified fidelity''
does not coincide with $\langle\psi_1|\hat\rho_{2}|\psi_1\rangle$,
but differs from it by the factor $[{\rm Tr}(\hat{\rho}_2^2)]^{-1/2}$.
Sometimes \cite{Joz,Schum} this is considered as a flaw.
But in our problem the factor discussed has real physical meaning, because it
permits us to distinguish mixed states from their pure ``partners''
possessing the same probability distribution functions of quanta
(see examples below).
Another difference between the cited papers and our approach
is that we consider ``fidelity'' not to a fixed state, but to the
whole family of ``classical'' states.

We shall use the notation $f(\eta,\alpha)$ for the function
${\rm Tr}(\hat{\rho}^{\prime}\hat{\rho}_c^{\prime})$,
whose maximum gives the value of classicality.
Sometimes, it can be difficult to find a maximum with respect to
both variables. Then one can calculate
partial maxima with respect to one variable, fixing zero value for
another. We shall denote such
``reduced classicalities'' as $\widetilde{\cal C}_{\alpha}$ or
$\widetilde{\cal C}_{\eta}$, where the subscript indicates the parameter,
with respect to which the maximum was calculated.

\subsection{Classicality of the Fock states}

For the Fock state $|n\rangle$, the quantity
${\rm Tr}(\hat{\rho}\hat{\rho}_c)$ is reduced to the probability
$\langle n|\hat{\rho}_c|n\rangle$ of discovering $n$ quanta in the
displaced thermal state. This probability was calculated by many authors,
e.g., in \cite{MolGla,Sem84,BiVour,DOM94}.
Using the results of \cite{DOM94} and
taking into account the expression for purity of the thermal state
(it does not depend on the shift parameter $\alpha$),
\be
\mbox{Tr}\left(\hat\rho_c^2\right)= \frac{1-\eta}{1+\eta},
\label{purth}
\ee
we arrive at the function
\beqn
f_n(\eta,\alpha)&=& \sqrt{1-\eta^2}\,\eta^n
\exp\left(-|\alpha|^2\sqrt{1-\eta^2}\right)
\nonumber \\ && \times
L_n\left(-|\alpha|^2(1-\eta)^2/\eta\right),
\label{fFdts}
\eeqn
where $L_n(z)$ is the Laguerre polynomial.
For $\eta=0$ we have the well-known Poissonian distribution of the
coherent state,
\[
f_n(0,\alpha)=\exp\left(-|\alpha|^2\right)|\alpha|^{2n}/n!,
\]
which has maximum at $|\alpha|^2=n$. Therefore, confining ourselves
to pure coherent reference states, we obtain for the reduced classicality
of the Fock state the formula (cf. \cite{dw2})
$
\widetilde{\cal C}_{\alpha}^{(n)}= e^{-n}n^n/n!$,
which results in
$\widetilde{\cal C}_{\alpha}^{(n)}\approx (2\pi n)^{-1/2}$ for $n\gg 1$.

On the other hand, the function $f_n(\eta,0)$ has maximum at
$\eta_n=\sqrt{n/(n+1)}$, so that
\be
\widetilde{\cal C}_{\eta}^{(n)}=\sqrt{n^n/(1+n)^{n+1}}\,.
\label{CFth}
\ee
One can verify that
$\widetilde{\cal C}_{\eta}^{(n)} > \widetilde{\cal C}_{\alpha}^{(n)}$
for any $n\ge 1$. In particular,
$\widetilde{\cal C}_{\eta}^{(1)}=1/2$ and
$\widetilde{\cal C}_{\alpha}^{(1)}=1/e$,
whereas for $n\gg 1$ we have
$\widetilde{\cal C}_{\eta}^{(n)}\approx (e n)^{-1/2}$.
Moreover, the function $f_n(\eta,\alpha)$ has {\em negative\/} derivative
with respect to $|\alpha|^2$ at the point $|\alpha|^2=0$ for any $\eta_n$.
Therefore, the maxima of $f_n(\eta,\alpha)$ at the points $(\eta_n,0)$
are global, whereas the maxima at the points $(0,n)$ are only local:
see Fig.~1.
This means that the classicality of the Fock state is given by
(\ref{CFth}), and that the pure state $|n\rangle$ is more close not
to the pure (displaced) coherent state with
$\overline{n}_{coh} \equiv |\alpha|^2=n$,
but to the mixed (undisplaced) thermal state with
\[
\overline{n}_{th} \equiv \eta/(1-\eta)
=\sqrt{n}\left(\sqrt{n}+\sqrt{n+1}\right)
\]
(i.e., $\overline{n}_{th}\approx 2n$ for $n \gg 1$) and
\[
\mbox{Tr}\left(\hat\rho_c^2\right)=\left(\sqrt{n}+\sqrt{n+1}\right)^{-2}.
\]

\subsection{Classicality of Gaussian states}

Generic Gaussian states are characterized \cite{DOM94} by two variances
of the quadrature components, $\s_q$ and $\s_p$, their covariance,
$\s_{pq}$, and two displacement parameters in the phase plane $qp$.
It is clear, however, that for the fixed (co)variances, the quantity
${\rm Tr}(\hat{\rho}\hat{\rho}_c)$ is maximal for coinciding
displacement parameters of $\hat{\rho}$ and $\hat{\rho}_c$, which means
that it is sufficient
to consider the states with zero displacements, i.e.,
to find the maximum of the
function of single variable $F(\eta)\equiv f(\eta,0)$. Since
the statistical operator of the thermal state is diagonal in the Fock basis,
the function $F(\eta)$ can be easily expressed in terms of the
{\em generating function\/}
\[
G(z)\equiv \sum_{n=0}^{\infty} p_n z^n
\]
for the photon distribution $p_n\equiv\langle n|\hat\rho|n\rangle$:
\be
F(\eta) =\mu^{-1/2}\sqrt{1-\eta^2}\,G(\eta), \quad
\mu \equiv {\rm Tr}(\hat{\rho}^2).
\label{Feta}
\ee
The function $G(z)$ for generic Gaussian states was calculated by means
of different methods, e.g., in \cite{Sem84,DOM94,Chat89}. Using the form
given in \cite{DOM94}, we arrive at the function
\be
F(\eta)=2\left[\frac{\mu(1-\eta^2)}{a -2b \eta
+ c \eta^2}\right]^{1/2},
\label{FGa}
\ee
where
\[
a= 1 + \mu^2+2T\mu^2 , \quad b= 1-\mu^2, \quad c= 1+ \mu^2 -2T\mu^2 ,
\]
\[
\mu^2=\left[4\left(\s_p\s_q -\s_{pq}^2\right)\right]^{-1},
\quad T= \s_p +\s_q\,.
\]
One can easily verify that the maximum of function (\ref{FGa}) is
attained for $\eta_*=(1-\mu)/(1+\mu)$, which means that the thermal state
which is most close to the given Gaussian state has the same purity:
$\mu_c=\mu$. The classicality of the Gaussian states is given by a
simple formula
\be
{\cal C}^{(G)}=\sqrt{\frac{2}{1+\mu T}}\,.
\label{CG}
\ee
Note that $\mu T \ge 1$ for Gaussian states, and
the equality holds for thermal states.
The parameter $T$ is related to the mean number of quanta $\overline{n}$
in the unshifted Gaussian state as $T=1+2\overline{n}$. Therefore,
the classicality of
{\em pure\/} unshifted Gaussian states (which are nothing but
{\em squeezed vacuum states\/}) equals
${\cal C}^{(sqv)}=(\overline{n}+1)^{-1/2}$.
If $\mu<1$, then the minimal mean number of quanta in the Gaussian state
equals $\overline{n}_{min}=(1-\mu)/(2\mu)$. However, mixed Gaussian
states remain {\em unsqueezed\/} (that means, no one quadrature component
can have the variance less than the vacuum state value $1/2$ in
dimensionless units), if
$\overline{n}<\overline{n}_{c}=(1-\mu^2)/(2\mu^2)$.
Such states possess positive $P$-distributions \cite{MolGla}, and it was
proposed in \cite{MarMar02} to extend the family of ``classical'' states,
including all mixed unsqueezed Gaussian states. We suppose to
study such an ``extended classicality'' in another paper.

There exist
{\em pure\/} states possessing the same photon distribution function as
thermal ones; usually they are called {\em coherent phase states\/}
\cite{Ler,Wun01}:
\be
|\vep\rangle= \sqrt{1\!-\!|\vep|^2} \sum_{n=0}^{\infty} \vep^n
\:|n\rangle.
\label{vep}
\ee
Although the state (\ref{vep}) is most close
(at least with respect to the reduced $\widetilde{\cal C}_{\eta}$-measure)
to the thermal state
with $\eta=|\vep|^2$, its $\eta$-classicality is less than $1$
(precisely because we use renormalized operators in the definition
(\ref{def-C}), the function $F(\eta)$ in Eq. (\ref{Feta}) contains
the factor $\mu^{-1/2}$, but this factor equals unity now, as soon as
$\hat\rho_{\vep}^{\prime}=\hat\rho_{\vep}$):
\[
\widetilde{\cal C}^{(\vep)}_{\eta}=\sqrt{\frac{1-|\vep|^2}{1+ |\vep|^2}}
=\left(1+2\overline{n}\right)^{-1/2}\,,
\]
in accordance with our feeling that pure states, in general, are less
classical than their mixed partners.
For highly excited states
with the same mean value $\overline{n} \gg 1 $ we obtain the asymptotical
relations
\[
{\cal C}^{(sqv)}\approx \sqrt2\, \widetilde{\cal C}^{(\vep)}_{\eta}
\approx \sqrt{e}\,{\cal C}^{(Fock)}.
\]

\section{Anticlassicality measures}

As soon as the Fock states are usually considered as
``the most quantum'' states, we can use them as reference states
for the definition of the degree of ``anticlassicality'':
\be
{\cal A}=
\max_{n}\langle n|\hat\rho |n\rangle.
\label{def-A}
\ee
Perhaps, it could be worth including the displaced Fock states,
$|\tilde{n}\rangle \equiv \hat{D}(\alpha)|n\rangle$,
into the set of reference states, but here we shall not do it.
Actually, we can define two measures: one given by (\ref{def-A}),
where integer $n$ runs over all integers, {\em including $n=0$\/},
and another (denoted as ${\cal A}_1$),
where the vacuum state $|0\rangle$
(which is distinguished from all other states)
is excluded from the set of reference states.
To see the difference, let us consider an example of
coherent states $|\alpha\rangle$. One can easily find that
the dependence of ${\cal A}$ on the mean photon number
$\overline{n}=|\alpha|^2$ has different analytical forms in the
intervals $k\le \overline{n} \le k+1$, $k=0,1,\ldots\,$:
\be
{\cal A}^{(\alpha)}(\overline{n})=
\exp\left(-\overline{n}\right)\overline{n}^{k}/k!\,, \quad
k\le \overline{n} \le k+1.
\label{anticoh}
\ee
The function given by (\ref{anticoh}) is continuous,
but it has jumps of derivatives
at the points $\overline{n}_k=k$: the right derivatives at these
points are equal to zero, while the left derivatives have finite
negative values.
However, ${\cal A}^{(\alpha)}(\overline{n})$ goes to $1$ as
$\overline{n}\to 0$, and this may seem strange.

The function ${\cal A}_1^{(\alpha)}(\overline{n})$ coincides with
${\cal A}^{(\alpha)}(\overline{n})$ for $\overline{n} \ge 1$. But for
$\overline{n}<1$ its behaviour is quite different:
\[
{\cal A}_1^{(\alpha)}(\overline{n})=
\overline{n}\,\exp\left(-\overline{n}\right), \quad 0\le \overline{n} \le 1.
\]
The dependence  ${\cal A}_1^{(\alpha)}(\overline{n})$ is shown in Fig.~2.
Asymptotically, for $\overline{n} \gg 1$, we have
(using the Stirling formula)
${\cal A}^{(\alpha)}_1 \approx (2\pi \overline{n})^{-1/2}$.

As another example we consider the squeezed
vacuum state. It is known that the photon distribution function
$p_n$ of this state is different from zero only for even values
$n=2m$, and it can be expressed in terms of the mean number of photons as
\cite{DOM94}
\be
p_{2m}^{(sqv)}=(1+\overline{n})^{-1/2}\frac{(2m)!}{(2^m\,m!)^2}
\left(\frac{\overline{n}}{1+\overline{n}}\right)^m.
\label{pmsq}
\ee
The sequence $p_{2m}$ decreases monotonously. Therefore we obtain a
monotonous dependence
\[
{\cal A}^{(sqv)}(\overline{n})=(1+\overline{n})^{-1/2},
\]
whereas the function
\[
{\cal A}_1^{(sqv)}(\overline{n})=
\frac{\overline{n}}{2(1+\overline{n})^{3/2}}
\]
has maximum ${\cal A}_{1max}^{(sqv)}=1/(3\sqrt3)$ for
$\overline{n}=2$ (and the most close Fock state is $|2\rangle$).

For the coherent phase state (\ref{vep}) we obtain
\[
{\cal A}^{(\vep)}(\overline{n})=(1+\overline{n})^{-1}, \quad
{\cal A}_1^{(\vep)}(\overline{n})=\overline{n}(1+\overline{n})^{-2},
\]
with ${\cal A}_{1max}^{(\vep)}=1/4$ at $\overline{n}=1$.
According to definition (\ref{def-A}), the same results must hold for
thermal states, since they have the same photon distribution function
$\langle n|\hat\rho |n\rangle$. However, we can distinguish
pure and mixed states
having the same diagonal elements in the Fock basis, playing again with
the purity parameter $\mu=\mbox{Tr}(\hat\rho^2)$. But in contrast
to the case of classicality parameter, now we should, sooner,
{\em multiply\/} the quantity $\langle n|\hat\rho |n\rangle$ by $\mu$
(or some power of $\mu$, if one wishes), rather than divide it by $\mu$
(or $\sqrt\mu$), because the intuition tells us that mixed states are
more anticlassical than pure ones.
Adopting the definitions
\be
\widetilde{\cal A}= \mu\,\max_{n}\langle n|\hat\rho |n\rangle, \quad
\widetilde{\cal A}_1= \mu\,\max_{n\ge 1}\langle n|\hat\rho |n\rangle,
\label{def-Atil}
\ee
we obtain for the thermal states the expressions
\[
\widetilde{\cal A}^{(th)}(\overline{n})=
\left[(1+\overline{n})(1+2\overline{n})\right]^{-1},
\]
\[
\widetilde{\cal A}_1^{(th)}(\overline{n})=
\frac{\overline{n}}{(1+\overline{n})^2(1+2\overline{n})}.
\]
These three cases are illustrated in Fig.~3.
The asymptotical dependences at $\overline{n} \gg 1$ are as follows:
\[
{\cal A}_1^{(sqv)}\sim \frac1{2\sqrt{\overline{n}}}, \quad
{\cal A}_1^{(\vep)}\sim \frac1{2\overline{n}}, \quad
{\cal A}_1^{(th)}\sim \frac1{2\overline{n}^{2}}.
\]

For generic (mixed and shifted) Gaussian states, the dependences
${\cal A}_1(\overline{n})$ and $\widetilde{\cal A}_1(\overline{n})$
may be more complicated, especially in the regime of strong irregular
oscillations of the photon distribution function $p_n$
\cite{osc,DKMosc}. We reserve this case for another study.

\section{Conclusion}

We have introduced new measures of ``classicality'' and ``anticlassicality''
of quantum states, which characterize the ``fidelities'' of the chosen
state to the whole families of states considered as ``classical'' or
``maximally quantum''. Due to simplicity, these measures
can be calculated analytically for many important sets of quantum states,
thus providing additional information on their properties.
In particular, we have shown that the most close to arbitrary Gaussian states
are thermal states with the same degree of quantum mixing (purity).
At the same time, there are (mixed) thermal states which are more close to
the Fock states than (pure) coherent states.
As soon as the new measures distinguish coherent quantum superpositions
(``cat states'') from quantum mixtures,
they could be used in studies of decoherence processes.
Also, it would be interesting to establish relations between the new and
existing measures of classicality, such as Mandel's parameter,
``nonclassical depth'' \cite{Lee91}, and so on, or to try to ``order''
different families of quantum states in accordance with their degrees of
(anti)classicality. We hope to report on the results
of studies in these directions somewhere.

\section*{Acknowledgments}

VVD and MBR acknowledge support of the Brazilian agencies CNPq and
FAPESP, respectively.


\newpage

\begin{figure}
\caption{The renormalized Hilbert--Schmidt fidelity between the Fock
state $|3\rangle$ and displaced thermal states.}
\label{F1}
\end{figure}

\begin{figure}
\caption{The anticlassicality measure ${\cal A}_1^{(\alpha)}$
of coherent states versus the mean number of quanta $\overline{n}$.}
\label{F2}
\end{figure}

\begin{figure}
\caption{The anticlassicality measures ${\cal A}_1(\overline{n})$
for the squeezed vacuum and phase coherent states, compared with
the modified measure
$\widetilde{\cal A}_1(\overline{n})$ for the thermal states.}
\label{F3}
\end{figure}

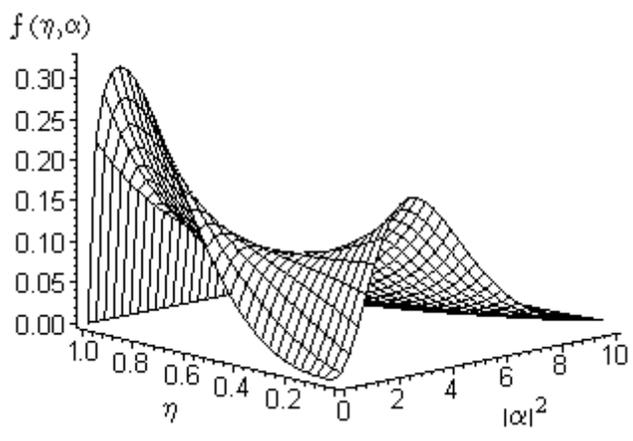
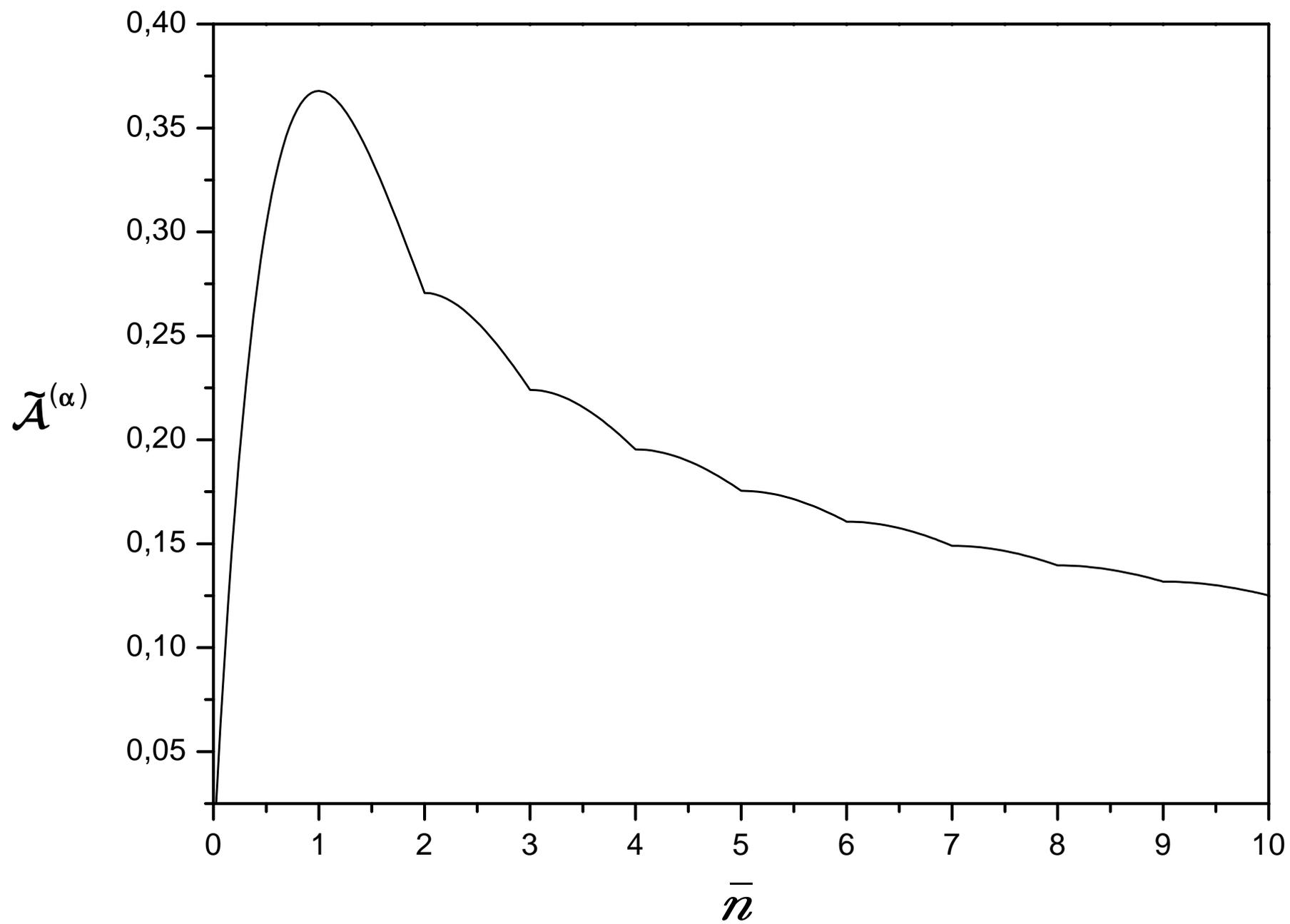
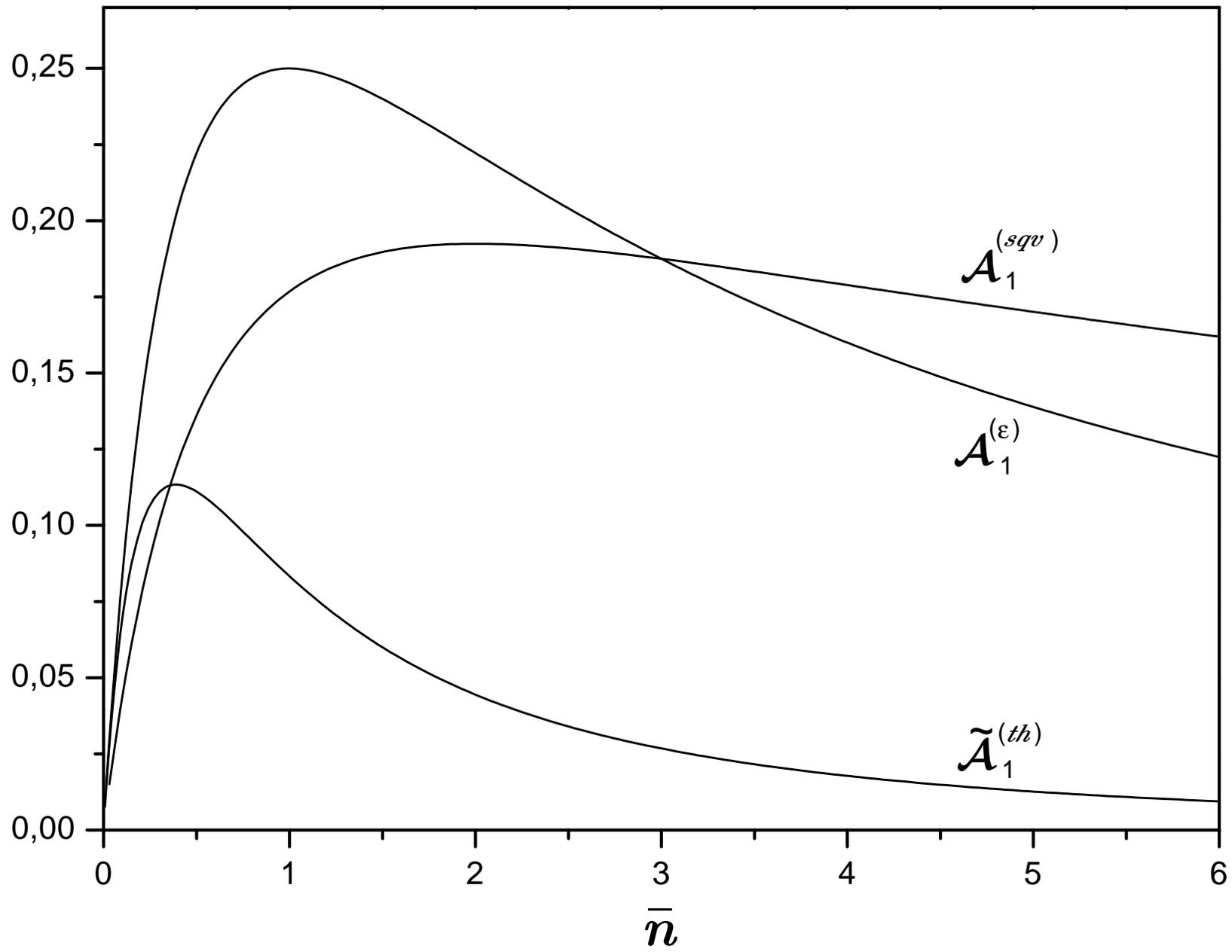
\end{document}